\documentclass[fleqn,10pt]{wlscirep}
\usepackage[utf8]{inputenc}
\usepackage[T1]{fontenc}
\usepackage[version=4]{mhchem}
\usepackage{float}
\usepackage{upgreek}

\title{Dual wavelength brillouin laser terahertz source stabilized to carbonyl sulfide rotational transition}

\author[1,*]{James Greenberg}
\author[1]{Brendan M. Heffernan}
\author[1]{William F. McGrew}
\author[1]{Keisuke Nose}
\author[1]{Antoine Rolland}
\affil[1]{Boulder Research Labs, IMRA America, Inc., 1551 S Sunset St. Suite C, Longmont CO, 80501}
\affil[*]{jgreenbe@imra.com}

\begin{abstract}
Optical-based terahertz sources are important for many burgeoning scientific and technological applications. Among such applications is precision spectroscopy of molecules, which exhibit rotational transitions at terahertz frequencies. Stemming from precision spectroscopy is frequency discrimination and stabilization of terahertz sources. Because many molecular species exist in the gas phase at room temperature, their transitions are prime candidates for practical terahertz frequency references. We demonstrate the stabilization of a low phase-noise, dual-wavelength Brillouin laser (DWBL) terahertz oscillator to a rotational transition of carbonyl sulfide (\ce{OCS}). We achieve an instability of $1.2\times10^{-12}/\sqrt{\tau}$, where $\tau$ is the averaging time in seconds. The signal-to-noise ratio and intermodulation limitations of the experiment are also discussed. We thus demonstrate a highly stable and spectrally pure terahertz frequency source. Our presented architecture will likely benefit metrology, spectroscopy, precision terahertz studies, and beyond.
\end{abstract}
\begin{document}

\flushbottom
\maketitle
\thispagestyle{empty}

\section*{Introduction}

The terahertz (0.1–10 THz) region of the electromagnetic spectrum has garnered significant interest due to its potential to address a diverse range of scientific and technological challenges.\cite{leitenstorfer2023} Higher bandwidth requirements and an ever-growing number of internet connected devices are creating spectral congestion in currently available wireless communication bands. So-called ``beyond 5G'' wireless communications will require access to higher carrier frequencies, where recent advances have demonstrated the utility of compact terahertz sources enabling high-speed data transfer.\cite{heffernan2023} High-frequency radar, which has been used to detect small foreign objects on aircraft runways,\cite{futatsumori2022} can improve the spatial resolution of traditional radar with less vulnerability to inclement weather as lidar. Future developments in radioastronomy will rely on high-frequency oscillators with low timing jitter to push the spatial resolution limits of terahertz observations from space.\cite{carpenter2020} Furthermore, laboratory-based precision spectroscopy uses narrow-line width sources to establish frequency benchmarks for the remote identification of molecules.\cite{aiello2022, djevahirdjian2023} Despite these applications, a ``terahertz gap'' persists, defined by the absence of mature technologies that deliver simultaneous spectral purity and stability at these frequencies.\cite{consolino2017}

One widely used approach to generating stable terahertz radiation involves the frequency multiplication of a microwave synthesizer, typically referenced to an atomic standard.\cite{maiwald2003} While this method transfers the fractional frequency stability to higher frequencies, it also multiplies the synthesizer’s phase noise, thereby degrading spectral purity.\cite{rubiola2009} This trade-off between purity and stability has led to the investigation of alternative terahertz sources based on down-conversion from optical frequencies involving division and/or photomixing. Optical sources have separately demonstrated both high spectral purity \cite{freeman2017, tetsumoto2021a} and the ability to transfer stability from microwave atomic standards.\cite{zhang2019a}

A key advancement in optical-based terahertz sources is the dual-wavelength Brillouin laser (DWBL): a tunable source spanning frequencies from 300\,GHz to 3\,THz with unmatched spectral purity.\cite{heffernan2024} The DWBL has already proven its high performance in wireless communication systems at 300 GHz, reaching a data transfer rate of 200\,Gbit/s over a distance of 200\,m.\cite{maekawa2024} While the DWBL exhibits low phase noise at high Fourier frequencies, it undergoes a frequency drift of several hundred kHz over the course of a day. Stabilization could be achieved by referencing microwave or optical sources through complex optical frequency comb architectures\cite{shin2023}. An alternative, streamlined approach, would be the development of a native terahertz reference, such as terahertz cavities~\cite{braakman2011,alligooddeprince2013} or utilizing molecular rotations as frequency discriminators (discussed below). With its inherent spectral purity, the DWBL is particularly well-suited for probing native terahertz frequency references. This is due to the intermodulation effect, which limits how well a source can be stabilized, by its phase noise \cite{audoin1991,bahoura2003}. These references, in-turn, could effectively mitigate the frequency drift and enhance the long-term stability of the laser.

Many small gas-phase molecules, such as \ce{N2O}, \ce{OCS}, and \ce{HCN}, exhibit well-defined rotational transitions at room temperature, which overlap with the DWBL’s tuning range.\cite{townes1975} These rotational transitions provide stable, absolute frequency references and can be probed using absorption spectroscopy in compact path lengths. Unlike atomic transitions, molecular rotational transitions form a ladder of nearly equidistant steps in frequency space, offering multiple terahertz-frequency references from a single molecule, which may be advantageous for future wireless communication channels\cite{dang2020, schneider2012} and their aggregation.\cite{ducournau2023} Using these molecular transitions circumvents the need for frequency translation from microwave or optical references, providing a multitude of transition frequencies with varying sensitivities to environmental perturbations. These characteristics enable the practical realization of a high-purity, low-drift terahertz source.

Stabilizing a terahertz source using molecular rotational transitions also represents an important step toward establishing a secondary frequency standard in the terahertz domain. Recent advances in this field have focused on reducing fractional frequency instability,\cite{greenberg2024} extending operations to higher frequencies,\cite{voigt2023} and achieving compact form factors \cite{wang2018}. However, the fundamental stability limits of rotational molecular transitions remain unclear, with the spectral purity of the free-running terahertz source being a key limiting factor (intermodulation limited).\cite{kim2022} The DWBL, with its superior spectral purity, offers a solution to these limitations, providing new insights into terahertz stabilization via molecular frequency references.

In this paper, we present an architecture that exploits the spectral purity of the DWBL in the context of molecular rotational spectroscopy. We demonstrate frequency discrimination and feedback, resulting in a stable terahertz wave, and we measure the instability of the radiation produced. Finally, we discuss the remaining stability limitations and prospects for future improvements and applications. Thereby, we have achieved a source simultaneously capable of producing a stable and pure terahertz wave making substantial progress towards bridging the terahertz gap.

\begin{figure}[t]
\centering
\includegraphics[width=\linewidth]{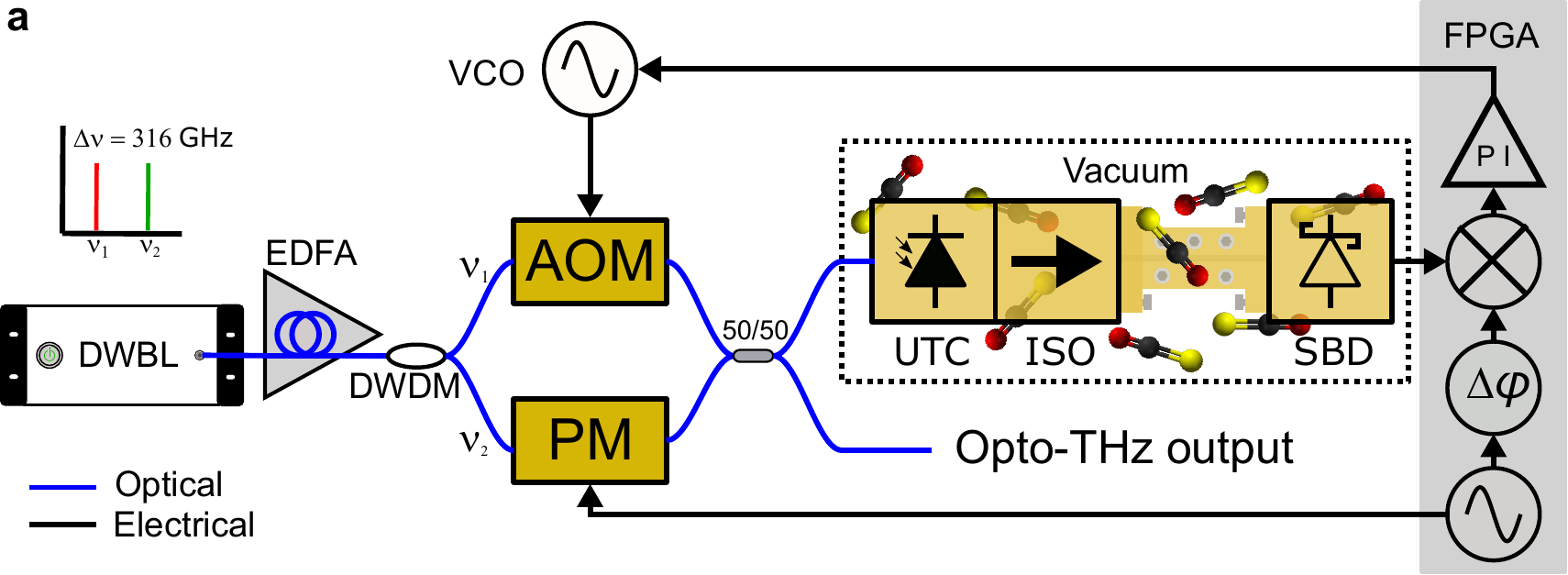}
\caption{Schematic representation of stabilized DWBL architecture. All fiber connections and components are single-mode, polarization maintaining fiber. See text for functional details. DWBL: dual-wavelength Brillouin laser; EDFA: erbium-doped fiber amplifier; DWDM: dense wavelength division multiplexer; VCO: voltage-controlled oscillator; AOM: acousto-optic modulator; PM: phase modulator; UTC: Uni-travelling-carrier photodiode; ISO: waveguide isolator; SBD: Schottky barrier diode; FPGA: field programmable gate array.}
\label{fig:setup}
\end{figure}

\section*{Results}
\subsection*{Phase modulation spectroscopy}

To realize frequency discrimination with the rotational state of a molecule, we performed phase modulation spectroscopy. A similar method has previously been described for a different oscillator and molecule.\cite{greenberg2024} In this report, we describe the oscillator architecture required to utilize the DWBL as a terahertz source, which is shown schematically in Fig. \ref{fig:setup}. A complete experimental diagram is also provided in the Supplementary Information (SI) Fig. S1\textbf{a}. The native output of the DWBL was two optical tones at a user-defined frequency difference in a single optical fiber. Since fast phase modulators (bandwidth $>1$\,kHz) do not yet exist in the terahertz band, a single optical tone was modulated instead ($f_{mod}=2$\,MHz). The modulation was inherited by the terahertz wave via photomixing. A dense wavelength division multiplexer (DWDM) was used to separate the two optical tones, one of which went through an electro-optic phase modulator (PM), before being recombined and photomixed by a uni-traveling carrier photodiode (UTC-PD). The other optical tone passed through an acousto-optic modulator (AOM) for fast frequency shifting. This allowed for fine-tuning of the terahertz frequency within a narrow frequency range, limited by the AOM bandwidth ($\pm 5$, MHz). The AOM was also used for frequency stabilization feedback from the frequency discriminator, as discussed in the next section.

To probe the rotational transitions of the \ce{OCS} molecule, a compact waveguide spectrometer was built and placed inside a rough vacuum chamber (base pressure $\approx 4$\, mTorr) pictured in the SI, Fig. S1\textbf{b}. The optically carried terahertz wave, and accompanying modulation, was delivered to the UTC-PD via a vacuum fiber feedthrough. In the vacuum chamber, terahertz radiation ($60\,\upmu$W) was generated via photomixing of the DWBL by the UTC-PD. The terahertz radiation passed through an isolator, followed by  13\,cm of rectangular waveguide, and was detected by a zero-bias Schottky barrier diode (SBD). Similarly, the electrical signal from the SBD was delivered to signal processing electronics via a coaxial (SMA) vacuum electrical feedthrough. We adopted this architecture to minimize terahertz etalons/standing waves and collision-induced broadening from gases at atmospheric pressure.

\begin{figure}[t]
\centering
\includegraphics[width=\linewidth]{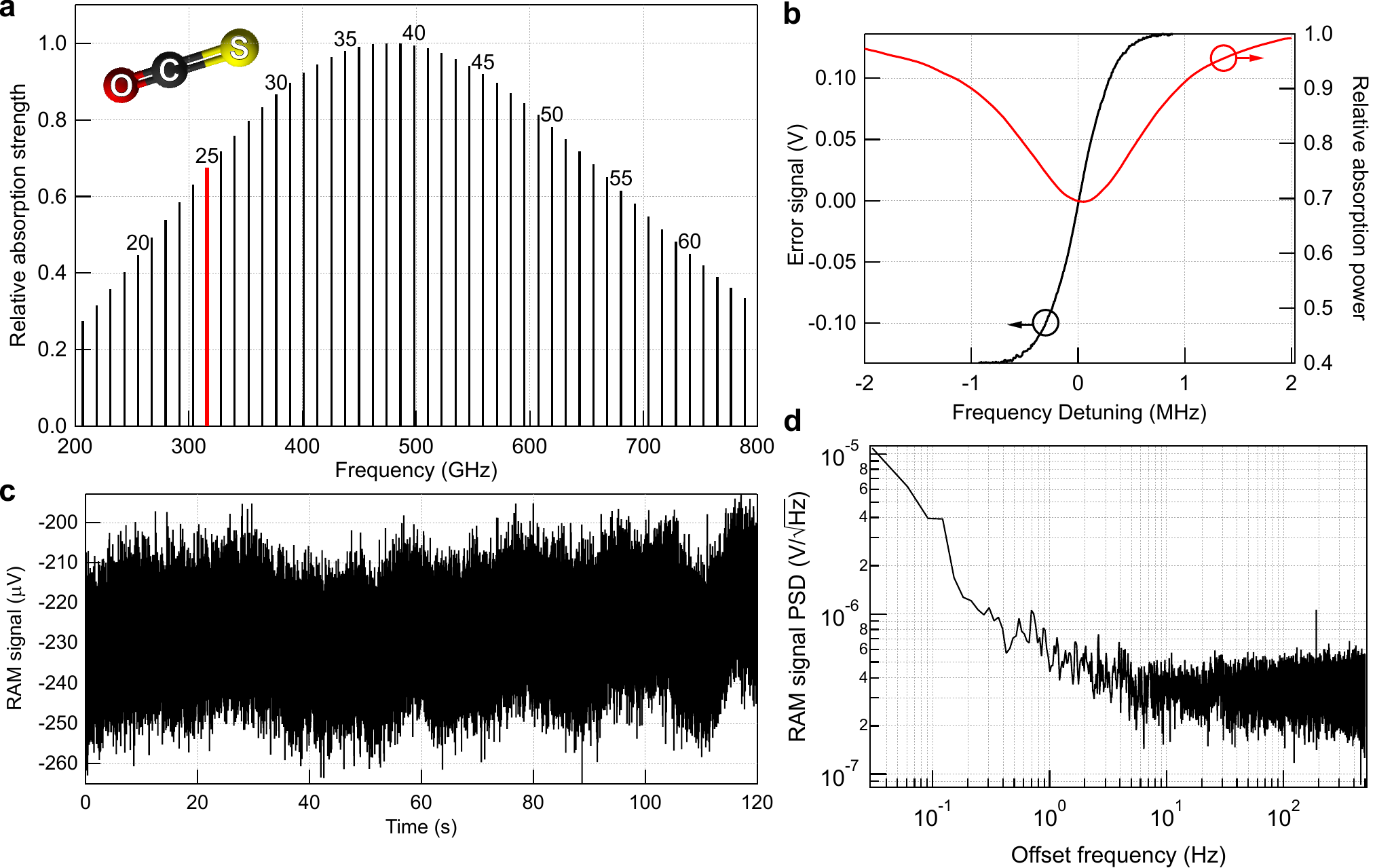}
\caption{Rotational ladder of \ce{OCS} with signal and noise floor measurements. \textbf{a:} Plot of rotational transitions at room temperature, normalized to the strongest absorption, as a function of terahertz radiation frequency. The numbers above the transitions are the lower rotational state quantum number, J. \textbf{b:} Error and absorption signals corresponding to  typical experimental conditions. The error signal axis is on the left. The absorption signal axis is on the right, normalized to the power detected without gas present, showing a peak absorption strength of 30\%. \textbf{c:} Time trace of the signal detected without gas. The long term fluctuations are a direct measurement of the residual amplitude modulation (RAM). \textbf{d:} Power spectral density (PSD) of the RAM signal time trace. These data are indicative of the spectrometer noise floor.}
\label{fig:snr}
\end{figure}

A small quantity of \ce{OCS} (70\,mTorr) was admitted into the vacuum chamber, and this gas diffused into the terahertz waveguide, where the gas molecules interacted with the terahertz radiation. Specifically, the terahertz wave was tuned to be resonant with the $J'=26 \leftarrow J''=25$ transition at $\nu_0 = 316.146$, GHz, which is highlighted in Fig. \ref{fig:snr}\textbf{a}. Scanning the terahertz frequency detuning around the resonance produced the measured absorption profile shown in  Fig. \ref{fig:snr}\textbf{b}. The pressure broadened profile was Lorentzian with a half-width at half-maximum of $\Gamma \approx 700$\,kHz. More details on the absorption measurement are provided in the SI (see Fig. S3).

The SBD detected amplitude variations at $f_{mod}$ from the dispersion of \ce{OCS} near resonance. FPGA-based signal processing electronics performed lock-in detection of the dispersion, generating an error signal, and thus achieved phase modulation spectroscopy. A truncated terahertz frequency scan, shown by the black line in Fig. \ref{fig:snr}\textbf{b}, measured the error signal from the \ce{OCS} resonance. The error signal, which corresponds to the derivative of the absorption resonance, realized the frequency discrimination needed for frequency stabilization.

The amplitude of the error signal ($V_{sig} = 140$\,mV) can be compared to the noise in the measurement to determine the signal-to-noise ratio (SNR) of the spectrometer. The noise floor in the experiment was measured by removing the \ce{OCS} gas and taking a time trace of the error signal. This included noise from all possible sources such as the SBD, signal processing chain, and residual amplitude modulation (RAM)\cite{gehrtz1985}. The measured trace is shown in Fig. \ref{fig:snr}\textbf{c}. The longer timescale variations in the error signal were due to fluctuations in the RAM. Although the in-phase component of RAM was actively canceled (see methods), other sources of RAM fluctuations still existed. These are detailed in the discussion section. Converting the RAM signal to a power spectral density (PSD) revealed the white noise limited floor. The PSD corresponding to the RAM trace in Fig. \ref{fig:snr}\textbf{c} is shown in Fig. \ref{fig:snr}\textbf{d}. This resulted in a noise PSD of $V_{noise} = 350\,\text{nV}/\sqrt{\text{Hz}}$. Combined with the error signal amplitude, we calculated the spectrometer SNR in a $f_{bw} = 1\,$Hz bandwidth with $\text{SNR} \equiv 20 \log_{10}\left(V_{sig}/(V_{noise}\sqrt{f_{bw}})\right) = 111$\,dB\cite{vanier1981}.

\subsection*{Frequency feedback and metrology}

\begin{figure}[t]
\centering
\includegraphics[width=\linewidth]{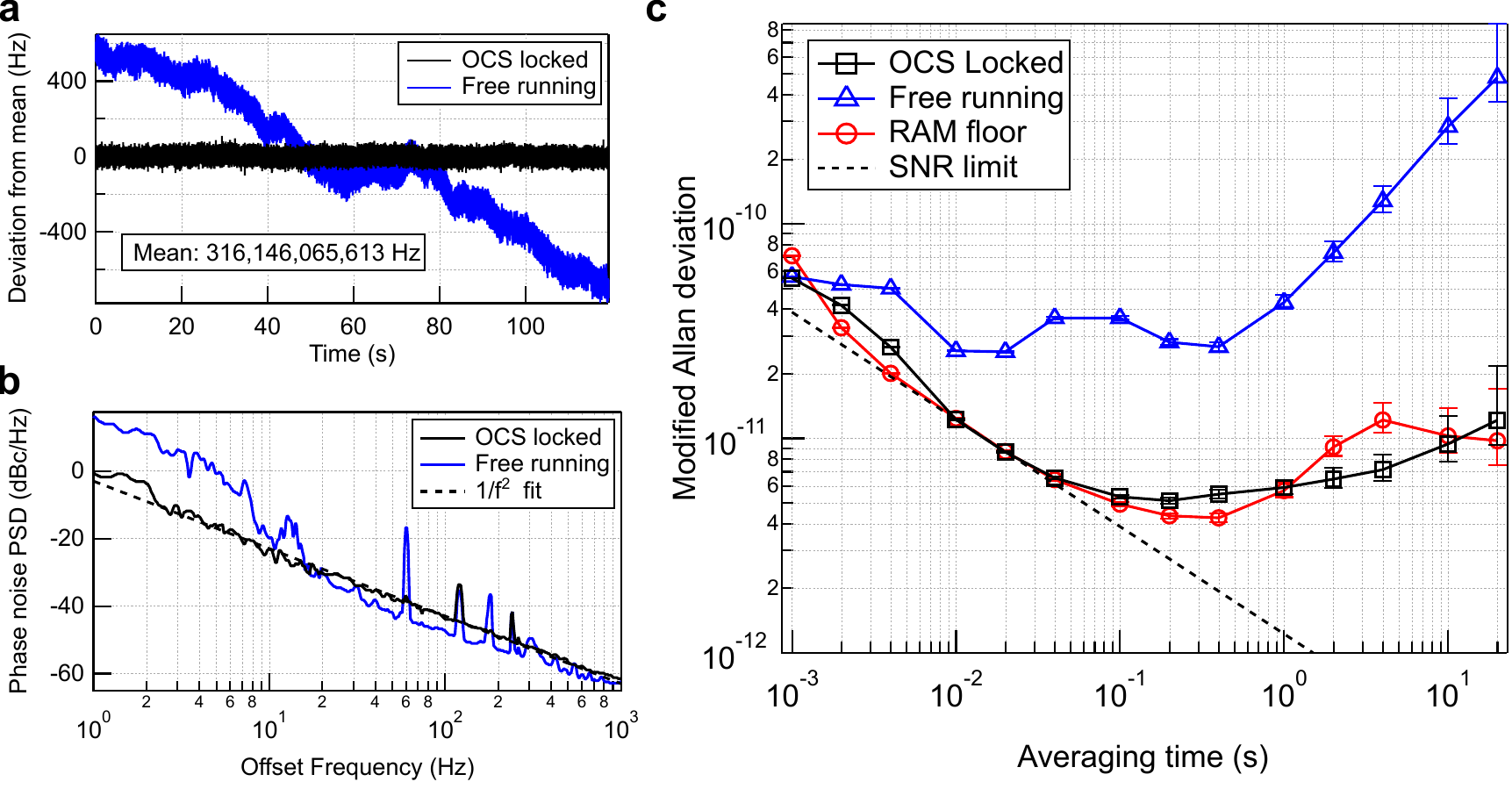}
\caption{Comparisons of free running and molecule-locked average frequencies, phase noises, and Modified Allan deviations. \textbf{a:} Time trace of frequency variations around the mean. \textbf{b:} Phase noise measured below 1\,kHz offset frequency. A 1/$f^2$ fit line is shown for reference. \textbf{c:} Modified allan deviations of fractional frequency fluctuations for the free running DWBL and locked to OCS data corresponding to the time traces in \textbf{a}. Also shown are scaled RAM fluctuations and SNR limit. These are described in the text. The error bars are 68.3\% confidence intervals.}
\label{fig:results}
\end{figure}

To cancel phase fluctuations and frequency drifts of the DWBL, we fed back the error signal to the VCO driving the AOM (see Fig. \ref{fig:setup}). Locking was achieved after the error signal was sent through a proportional-integrator (PI) filter with a corner frequency of $f_c=3$\,kHz. Once the DWBL was locked to the rotational transition, the terahertz frequency was stabilized and could be monitored for absolute frequency fluctuations. These fluctuations were used to evaluate the performance of the lock. We relied on an established technique for frequency down-conversion, which utilized an electro-optic (EO) comb, and counted the resultant RF signal that carried the fluctuations of the terahertz wave (see methods for more details).\cite{rolland2011} Fig. \ref{fig:results}\textbf{a} shows a comparison of the DWBL terahertz frequency fluctuations between free running and \ce{OCS} locked states. The overall drift of the DWBL locked to \ce{OCS}, at time scales greater than a few seconds, was reduced by more than an order of magnitude. 

Through the EO comb down-conversion, the source phase noise was also measured. Shown in Fig. \ref{fig:results}\textbf{b}, we observed a clear reduction in phase noise at low offset frequency $f<20\,$Hz. Additionally, the phase noise followed a slope of $f^-2$. This phase noise trend is characteristic of white frequency noise, as expected by an atomic, or in this case molecular, reference.\cite{rubiola2009} The PI feedback also added some noise to the free-running DWBL at intermediate $f$, motivating the choice of a relatively small $f_c = 3 \,$ kHz, so that the phase noise of the DWBL at $f>f_c$ was preserved. Fig. S4 in the SI shows an extended phase noise plot with measurement noise floor. 

To gain more insight into the nature of the long-term drift of the DWBL locked to \ce{OCS}, we calculated the modified Allan deviation of the fractional frequency deviations (data in Fig. \ref{fig:results}\textbf{a} divided by $\nu_0$), as shown in Fig. \ref{fig:results}\textbf{c}. The modified Allan deviation was chosen over the Allan deviation to remove white phase noise contributions made by the EO-comb down conversion to the frequency measurements and more accurately represent the stability of the terahertz wave, despite being measured optically. The data show that the fractional frequency of the stabilized terahertz oscillator was $1.2\times10^{-12}/\sqrt{\tau}$, where $\tau$ is the averaging time in seconds. The trend decreasing with $\sqrt{\tau}$ also revealed the characteristic of white frequency noise, in agreement with the phase noise data. 

However, the locked oscillator deviated from the trend after about 50\,ms, ultimately reaching a minimum fractional frequency instability of $5\times10^{-12}$. This deviation was not caused by the molecular frequency reference drifting, but rather by RAM. We also calculated the modified Allan deviation, $\sigma_{RAM}$, of the RAM data (shown in Fig. \ref{fig:snr}\textbf{c}) and scaled it to a frequency deviation by dividing by the peak slope of the error signal, $K = 460$\, V / GHz (shown in Fig. \ref{fig:snr}\textbf{b}).\cite{gillot2022} The considerable overlap between the locked OCS oscillator and the scaled RAM floor suggested that the frequency lock is limited by technical fluctuations in the error signal zero-crossing, which were caused by RAM as opposed to any fundamental physics of the \ce{OCS} transition.

\section*{Discussion}

To understand the limits of terahertz oscillator stability that can be achieved with molecular rotational transitions, we consider two possible limits to frequency stabilization. One fundamental limit to the clock instability in the presented architecture is the intermodulation limit arising from the phase noise properties of the terahertz local oscillator (here, the DWBL).\cite{audoin1991}. The intermodulation effect is caused by the mixing product of the free-running phase noise of the local oscillator at $2f_{mod}$ and the molecular error signal resulting from phase modulation spectroscopy. For the DWBL, the limit is estimated (in the quasi-static regime) to be below $3\times10^{-13}/\sqrt{\tau}$. Because the modulation frequency is large compared to the absorption linewidth ($f_{mod}\gg\Gamma$), there is a reduction in this limit by a factor of $(\Gamma/(2f_{mod}))^2\approx0.03$, \cite{bahoura2003} leading to an ultimate intermodulation limit below $1\times10^{-14}/\sqrt{\tau}$. While we did not achieve this limit, the DWBL is believed to be capable of supporting a terahertz wave with fractional frequency instability comparable to that of a hydrogen maser.

The practical limitation to the oscillator's stability in this study was the spectrometer SNR. We calculated an SNR limit for the fractional frequency,\cite{vanier1981,wang2018} to be $1.2\times10^{-12}/\sqrt{\tau}$. As seen graphically in Fig. \ref{fig:results}\textbf{c}, the calculated value agrees within the 68.3\% confidence interval of the measured data. This level of instability is comparable to research-grade commercial rubidium atomic clocks over the timescales investigated. Further optimization of the SNR to find the fundamental limit is a challenging engineering task with many interrelated variables. For example, optimal waveguide length must balance molecular absorption strength with waveguide power loss; for pressure and terahertz power, the absorption signal must be balanced with linewidth broadening mechanisms (intensity and collisions). Similarly, the detector introduces another set of parameters that include $1/f$ noise, shot noise, sensitivity, and dynamic range. Finally, significant improvements may be made through the use of coherent detection in the terahertz.\cite{siegel2006,deumer2024}

Despite SNR limitations, the present architecture has achieved an unprecedented level of terahertz fractional frequency stability with molecules at 100\,ms averaging time of $5\times10^{-12}$. However, the frequency instability does not average down for very long. Fluctuations in the RAM prevent the oscillator from achieving a lower ultimate instability. As stated in the results section, the RAM floor scales inversely with the max signal slope. This means that a stronger absorber, like hydrogen cyanide (\ce{HCN}), would allow a lower RAM floor. Similarly, using a higher frequency transition in \ce{OCS} would do the same, although technical limitations in commercially available components at higher terahertz frequencies may offset such an improvement. At the time of writing, the WR3.4 waveguide components provide the best performance for the highest frequency range (220--330\,GHz). Linewidth narrowing spectroscopic schemes, like saturated absorption (Lamb-dip) spectroscopy, could also improve the relative impact of RAM.\cite{winton1970,cazzoli2013}

Finding and mitigating the source of the RAM fluctuations is another route to reducing the instability. As described by several prominent studies on RAM,\cite{zhang2014,descampeaux2021,kedar2023} there are sources of RAM fluctuations that remain in this experiment. The quadrature component of the RAM error signal was too small to servo and thus was left uncontrolled. Unidentified etalons could have contributed, although optical isolators were employed to prevent this. Non-common fiber between the photodiode for detecting RAM and the UTC-PD may have allowed time-dependent polarization rotations that led to RAM fluctuations (see SI Fig. S1\textbf{a} for complete experimental diagram). Lastly, RAM can mediate the conversion of signal amplitude fluctuations into frequency fluctuations. Thus variations in the terahertz power and gas pressure, which otherwise should not affect the molecular frequency, can cause measured frequency shifts in the presence of a non-zero spectroscopic baseline.

Although our spectrometer was designed for frequency feedback, it was also capable of precision rotational spectroscopy. Here we report the mean transition frequency of $\nu0 = 316,146,065,613\pm3\,$Hz with a statistical uncertainty derived from the modified Allan deviation after 120\,s of measurement. We stress that we have not made any attempt to quantify systematic shifts in this experiment, which certainly contained an unknown offset, likely dominated by RAM. RAM may explain the 24\,kHz discrepancy with the most accurate value reported thus far, $316,146,089,000 \pm 1,200$, Hz.\cite{golubiatnikov2005} The 400 times improvement in precision realized in the present architecture underscores the ability of the DWBL and the spectrometer for rotational spectroscopy.

\section*{Conclusion and Outlook}

We have developed an architecture for stabilizing a DWBL to a molecular rotational transition. Using phase modulation spectroscopy, we demonstrated that \ce{OCS} can serve as a highly effective frequency discriminator at 316\,GHz. Through relatively slow feedback ($f_c=3$,kHz), the DWBL drift was suppressed to a fractional level of $1.2\times10^{-12}/\sqrt{\tau}$, while preserving unprecedented phase noise above $f_c$. As a result, we have produced the most stable terahertz wave referenced to a molecule to date.

The impacts of such achievements extend beyond precision rotational spectroscopy. Much like how optical communication systems had once used molecular transitions, such as acetylene or HCN, as frequency references for channel stabilization, this architecture opens the door for similar applications at terahertz frequencies. By leveraging molecular rotational transitions as highly precise frequency references, this technology could be applied to wireless communication systems for channelization and aggregation, enhancing spectral efficiency in the terahertz range. Such an approach would bring molecular precision to the task of managing and optimizing bandwidth across communication channels. Furthermore, this architecture establishes the groundwork for a terahertz frequency standard, enabling advancements in precision timekeeping and synchronization. Molecules are especially well-suited for this role due to their reduced sensitivity to temperature fluctuations and magnetic fields compared to atoms, making them ideal for use in demanding conditions.\cite{roslund2024} 

The discussion of instability limitations and the potential of this device for precision rotational spectroscopy highlights the need for further investigation into the SNR, which is influenced by molecular selection, experimental conditions, and detector characteristics. Ensuring a stable frequency reference will also require addressing the RAM observed in this work. Potential solutions include narrowing the linewidth through saturated absorption and integrating the system into a compact form factor to minimize fiber related RAM contributions. Nevertheless, the results presented here are already comparable to commercially available microwave Rb atomic clocks over the studied timescale, with significant potential for future advancements.

\section*{Methods}

\subsection*{RAM measurement and cancellation}
We actively mitigated RAM while the source was in continuously locked operation. This presented a challenge, as the optimum method to measure RAM required the \ce{OCS} to be removed from the spectrometer (as shown in Fig. \ref{fig:snr}\textbf{c}). Instead, we measured the RAM optically with a photodiode (RAM-PD), via a 99:1 coupler immediately before the UTC-PD. Fig. S1\textbf{a} in the SI shows a complete experimental diagram that includes the RAM feedback loop. Detecting the RAM optically allowed us to employ a strategy similar to that in other RAM cancellation studies\cite{zhang2014,gillot2022} An error signal was generated from lock-in detection of the RAM-PD at $f_{mod}$. This signal was then fed back to the DC phase of the PM. Additionally, the PM was temperature stabilized to minimize quadrature RAM drifts. In total, the RAM in our experiment was on the order of $10^{-6}$, in terms of fractional fluctuations. However, the cancellation degraded with time as some form of out-of-loop fluctuations were present in the experiment (see discussion).

\subsection*{EO comb down-conversion}
We performed frequency metrology on the terahertz wave indirectly in this study. The optically carried terahertz wave at the output (see Fig. \ref{fig:setup}) was down-converted into a radio frequency (RF) through an EO comb. \cite{rolland2011,greenberg2024}. The pair of optical tones was sent through a series of three phase modulators, driven at $f_{synth}=10.53,GHz$ with high power (30\,dBm) and adjustable relative phase. When the phases were tuned, over 15 side-bands were generated around each carrier, spaced by $f_{synth}$, and optically bridged the gap between the two tones. The EO comb spectrum then passed through an optical bandpass filter tuned to the region in the center of the two tones that contained overlapping sidebands. The filtered spectrum was photodetected where an RF beatnote $f_{beat}$ was produced. The relationship between the beat note and the terahertz frequency was then given by $f_{THz} = 2Nf_{synth} \pm f_{beat}$, where $N=15$ was the number of sidebands from one carrier to the center, filtered region, and the ambiguous sign was experimentally determined by adjusting $f_{synth}$ while keeping $f_{THz}$ fixed. The EO comb synthesizer was locked to a stable Rb reference, thereby removing any potential drifts of the synthesizer from the measurement.

The EO comb also transferred the phase noise of the terahertz wave to the RF via the above relationship, which allowed an RF phase noise analyzer to measure the terahertz phase noise. However, the measured phase noise also includes the synthesizer phase noise, scaled up by a factor of $2N$. This prevented the EO comb method from measuring the true phase noise of the DWBL at offset frequencies above 1\,kHz. See SI Fig. S4 for more details.

\section*{Data availability}
Raw data that support the findings of this study are available from the corresponding author upon reasonable request.

\section*{Author contributions statement}
J.G. and A.R. conceived the experiment. W.F.M. and K.N. made significant contributions to the experiment design. J.G. and B.M.H. built and conducted the experiment. J.G. analyzed the results and prepared the manuscript. All authors participated in interpretation of the results and reviewed the manuscript. A.R. supervised the project.

\section*{Competing interests}
J.G., B.M.H., and A.R. are inventors on a patent application that may in aspects be related but is not limited to the subject matter disclosed in the manuscript.

\end{document}


\flushbottom
\maketitle

\section{Full diagram and RAM compensation}

\begin{figure}[H]
\centering
\includegraphics[width=\linewidth]{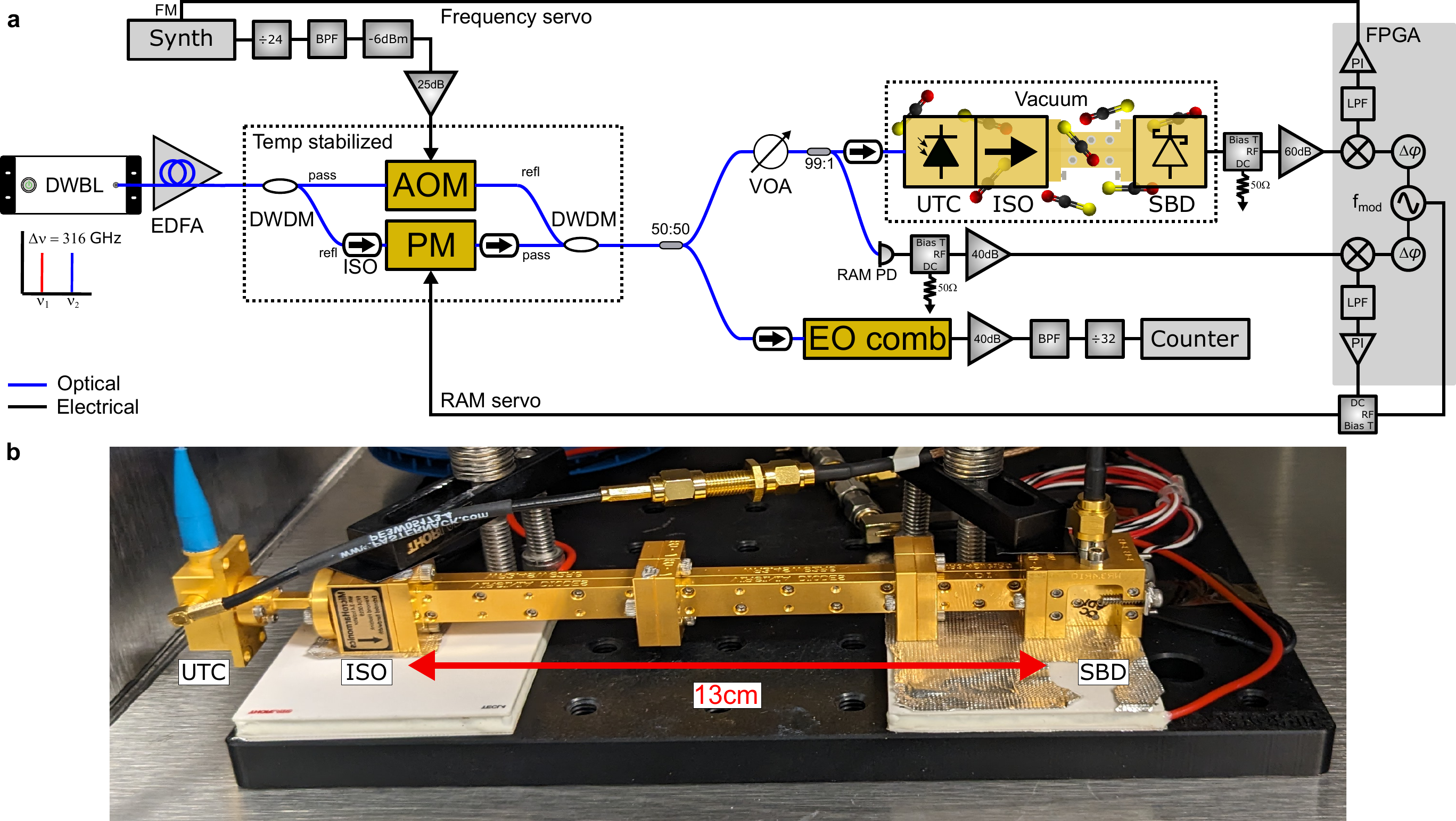}
\caption{Full experimental diagram with RAM feedback loop, and photo of in-vacuum spectrometer. \textbf{a:} DWBL: dual wavelength brillouin laser; EDFA: erbium-doped fiber amplifier; DWDM: dense wavelength division multiplexer; VCO: voltage-controlled oscilltor; AOM: acousto-optic modulator; PM: phase modulator; VOA: variable optical attenuator; PD: photodiode; UTC: Unitravelling-carrier photodiode; ISO: isolator; SBD: schottky barrier diode; FPGA: field programmable gate array; LPF: low-pass filter; PI: proportional-integrator filter; FM: frequency modulation. \textbf{b:} Photograph of waveguide spectrometer, with 13\,cm absorption path length, inside the vacuum chamber and waveguide components labelled. The pictured temperature stabilization system was not utilized in this experiment.}
\label{fig:diagram}
\end{figure}

Here we provide a full experimental diagram and description of components for reproducibility. The DWBL used in the experiment was built by us, mounted on a rack and described well in a previous publication.\cite{heffernan2024} The EDFA was also home-built, but was a standard dual-pump EDFA capable of 200\,mW output power (10\,mW input) and about 10dB noise figure. The placement and use of a single EDFA minimized any potential degradation of the optical signal-to-noise ratio of the DWBL ($\sim 65dB$ in a 0.02\,nm bandwidth). The two DWDMs were AFR PMDWDM-1-34-B-2-Q and PMDWDM-1-31-B-2-Q with 100GHz pass band, centered at 1550\,nm and 1552.5\,nm respectively. They were in opposing orientations to filter out the small amount of pass light that makes it through the reflect port. The optical isolators were also from AFR. The PM was ixblue MPX-LN-0.1, driven directly by the FPGA, liquid instruments Moku:Pro. The AOM, brimrose AMM-80-4-40-1550-2FP, was driven by a synthesizer with a DC-coupled analog input for frequency modulation. The synth, Agilent MXG n5181a, effectively acted as a VCO to drive the AOM. The output was amplified up to 25dBm by a mini-circuits ZHL-32a-S+. The phase noise considerations of this VCO, discussed below, justify the use of a frequency divider (Valon 3010a) to divide 1.92\,GHz down to 80MHz. An AFR 50:50 coupler split the opto-terahertz wave between the spectrometer and frequency readout. The spectrometer was made entirely of off-the-shelf parts: UTC (NTT IOD-PMJ-13001), waveguide isolator (microharmonics FR34M2), VDI WR3.4 waveguide straights and a zero-bias schottky detector (VDI WR3.4R10 ZBD-F06). The signal at $f_{mod}$ out of the SBD was amplified by a Pasternack 63014 before being processed by the Moku:Pro and then fed back to the VCO driving the AOM. Just before the UTC-PD, an optical signal was split by AFR 99:1 coupler to photodetect the RAM (Thorlabs pda10sc2). A Pasternack 63012 then amplified the signal before being processed by the Moku:Pro, and fed back to the PM. For frequency readout, the EO comb was built in-house, producing an RF frequency counted by a Microchip 53100A.

\section{VCO phase noise}

\begin{figure}[H]
\centering
\includegraphics[width=\linewidth]{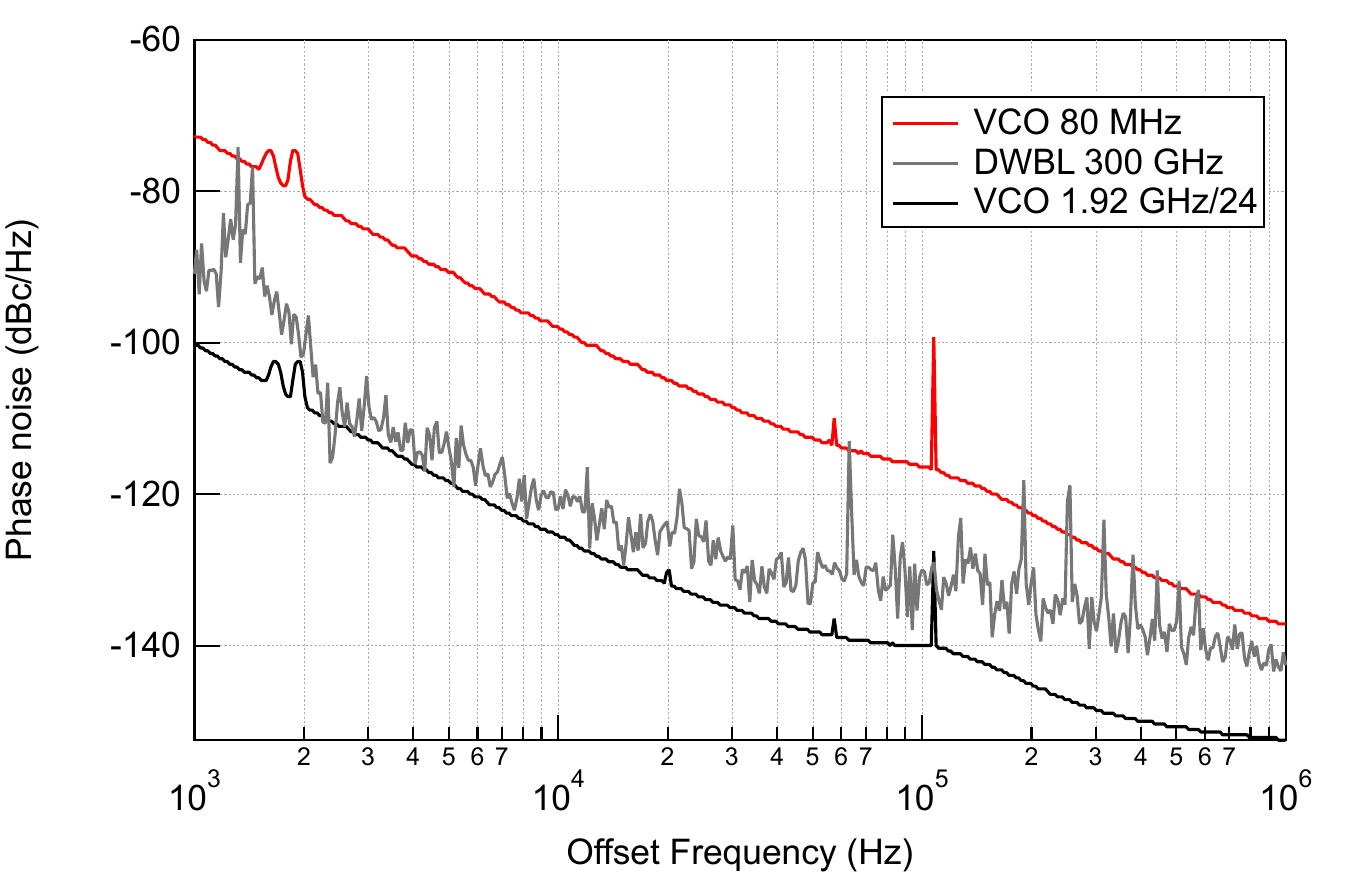}
\caption{Phase noise of VCO in two different configurations compared to free running DWBL. The DWBL data were reproduced with express permission from the authors.\cite{heffernan2024}}
\label{fig:vco}
\end{figure}

As mentioned in the previous section, we chose a higher-frequency VCO and divided down to drive the AOM for frequency shifting and feedback. Any phase noise on the AOM drive would be added directly to the terahertz phase noise because the two wavelengths of the DWBL were on separate paths. The Agilent MXG n5181a synthesizer generating the AOM drive happened to have very similar phase noise at both 80\,MHz and 2\,GHz outputs. Thus the phase noise of the needed 80MHz signal was decreased through frequency division of a higher input frequency. Fig. \ref{fig:vco} shows that this divided signal ensured that we were not adding any noise to the free-running DWBL. The converse is true for running at 80\,MHz directly and could potentially increase the intermodulation limit if used.

\section{Absorption measurement}

\begin{figure}[H]
\centering
\includegraphics[width=\linewidth]{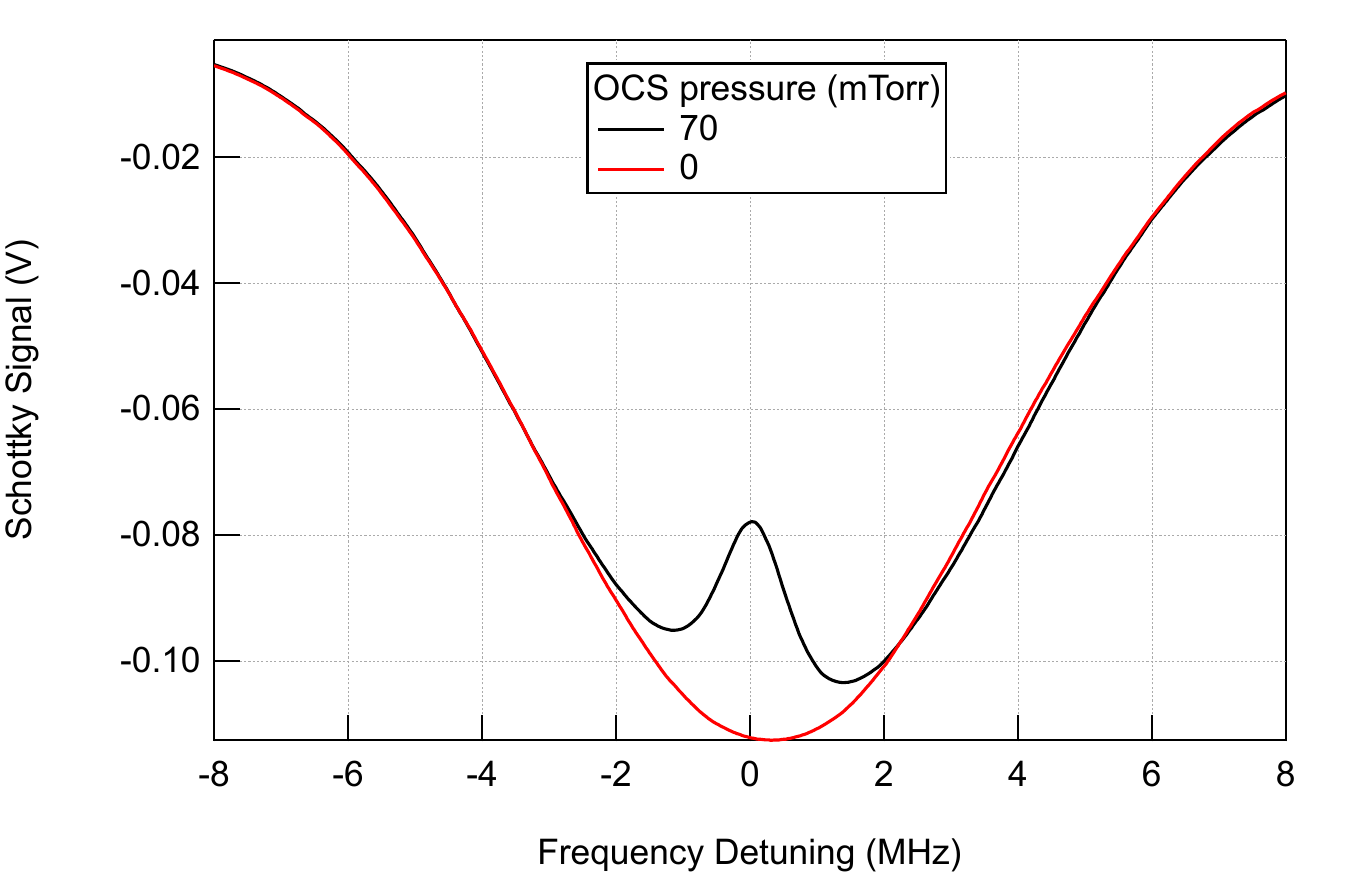}
\caption{Direct absorption and background measurements.}
\label{fig:abs}
\end{figure}

In the main experimental diagram (Fig. \ref{fig:diagram}) we split the RF and DC components generated by the SBD. In locked operation, we dumped the DC signal into a 50\,$\Omega$ terminator. However, that signal was proportional to the absorption strength. Thus, to determine the linewidth of the transition, one measured that DC signal as a function of DWBL frequency. Additionally, we used the undivided output of the VCO (mentioned in the previous section) to get a larger scan range. Fig. \ref{fig:abs} shows the DC signal from such a scan under normal experimental conditions (60\,$\mu$W emitted THz power). We disabled PM as RAM, and the $2f_{mod}$ sidebands could influence the resulting measured linewidth. The scan showed a large power variation vs frequency, which is the result of the AOM's narrow tuning range. Thus, to obtain an accurate measure of absortion, we performed the scan with and without gas and divided the two. The divided data and the resulting linewidth presented in the main text originate from the two scans shown here.

\section{Phase noise measured by EO comb}

\begin{figure}[H]
\centering
\includegraphics[width=\linewidth]{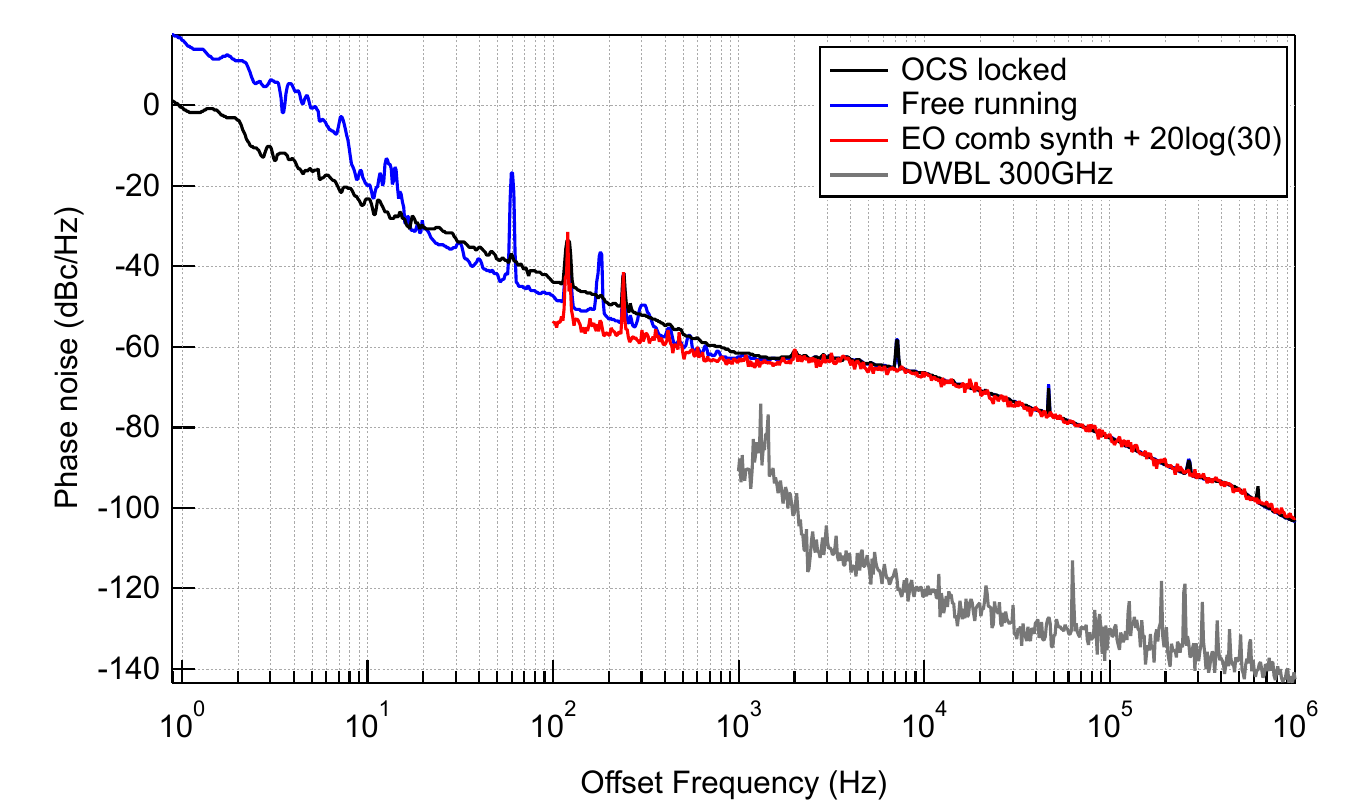}
\caption{Full phase noise measured by the EO comb. Shows synthesizer limited measurement above 1\,kHz and actual DWBL phase noise for comparison. The DWBL data were reproduced with express permission from the authors.\cite{heffernan2024}}
\label{fig:eocombpn}
\end{figure}

Fig. \ref{fig:eocombpn} shows the phase noise of the free running and locked DWBL, from the experiment, extended to higher offset frequency. Also included for reference are the synthesizer phase noise, scaled by a factor of $2N = 30$, and the true DWBL phase noise. The data show that the phase noise measurements were for sure limited by the EO comb above an offset frequency of 1\,kHz. It is possible that the free running data were also limited/influenced by the EO comb in the range of 20\,Hz to 1\,kHz. Regardless, there would be no impact on the interpretation of results presented in the main text.